\documentstyle[twoside,epsf]{article}

\catcode`\@=11
\long\def\@makefntext#1{
\protect\noindent \hbox to 3.2pt {\hskip-.9pt
$^{{\eightrm\@thefnmark}}$\hfil}#1\hfill}       

\def\@makefnmark{\hbox to 0pt{$^{\@thefnmark}$\hss}}    

\def\ps@myheadings{\let\@mkboth\@gobbletwo
\def\@oddhead{\hbox{}
\rightmark\hfil\eightrm\thepage}
\def\@oddfoot{}\def\@evenhead{\eightrm\thepage\hfil
\leftmark\hbox{}}\def\@evenfoot{}
\def\sectionmark##1{}\def\subsectionmark##1{}}

\oddsidemargin=\evensidemargin
\newcounter{sectionc}\newcounter{subsectionc}\newcounter{subsubsectionc}
\renewcommand{\section}[1] {\vspace{12pt}\addtocounter{sectionc}{1}
\setcounter{subsectionc}{0}\setcounter{subsubsectionc}{0}\noindent
    {\tenbf\thesectionc. #1}\par\vspace{5pt}}
\renewcommand{\subsection}[1] {\vspace{12pt}\addtocounter{subsectionc}{1}
\setcounter{subsubsectionc}{0}\noindent
{\bf\thesectionc.\thesubsectionc. {\kern1pt \bfit #1}}\par\vspace{5pt}}
\renewcommand{\subsubsection}[1] {\vspace{12pt}\addtocounter{subsubsectionc}{1}
    \noindent{\tenrm\thesectionc.\thesubsectionc.\thesubsubsectionc.
    {\kern1pt \tenit #1}}\par\vspace{5pt}}
\newcommand{\nonumsection}[1] {\vspace{12pt}\noindent{\tenbf #1}
    \par\vspace{5pt}}

\newcounter{appendixc}
\newcounter{subappendixc}[appendixc]
\newcounter{subsubappendixc}[subappendixc]
\renewcommand{\thesubappendixc}{\Alph{appendixc}.\arabic{subappendixc}}
\renewcommand{\thesubsubappendixc}
    {\Alph{appendixc}.\arabic{subappendixc}.\arabic{subsubappendixc}}

\renewcommand{\appendix}[1] {\vspace{12pt}
        \refstepcounter{appendixc}
        \setcounter{figure}{0}
        \setcounter{table}{0}
        \setcounter{lemma}{0}
        \setcounter{theorem}{0}
        \setcounter{corollary}{0}
        \setcounter{definition}{0}
        \setcounter{equation}{0}
        \renewcommand{\thefigure}{\Alph{appendixc}.\arabic{figure}}
        \renewcommand{\thetable}{\Alph{appendixc}.\arabic{table}}
        \renewcommand{\theappendixc}{\Alph{appendixc}}
        \renewcommand{\thelemma}{\Alph{appendixc}.\arabic{lemma}}
        \renewcommand{\thetheorem}{\Alph{appendixc}.\arabic{theorem}}
        \renewcommand{\thedefinition}{\Alph{appendixc}.\arabic{definition}}
        \renewcommand{\thecorollary}{\Alph{appendixc}.\arabic{corollary}}
        \renewcommand{\theequation}{\Alph{appendixc}.\arabic{equation}}
        \noindent{\tenbf Appendix \theappendixc #1}\par\vspace{5pt}}
\newcommand{\subappendix}[1] {\vspace{12pt}
        \refstepcounter{subappendixc}
        \noindent{\bf Appendix \thesubappendixc. {\kern1pt \bfit #1}}
    \par\vspace{5pt}}
\newcommand{\subsubappendix}[1] {\vspace{12pt}
        \refstepcounter{subsubappendixc}
        \noindent{\rm Appendix \thesubsubappendixc. {\kern1pt \tenit #1}}
    \par\vspace{5pt}}

\newcommand{\textlineskip}{\baselineskip=13pt}
\newcommand{\smalllineskip}{\baselineskip=10pt}

\newcommand{\copyrightheading}[1]
    {\vspace*{-2.5cm}\smalllineskip{\flushleft
     }}

\def\keywords#1{{
    \centering{\begin{minipage}{4.5in}\footnotesize\baselineskip=10pt
    {\footnotesize\it Keywords}\/: #1
     \end{minipage}}\par}}

\renewenvironment{thebibliography}[1]
        {\frenchspacing
     \ninerm\baselineskip=11pt
         \begin{list}{\arabic{enumi}.}
        {\usecounter{enumi}\setlength{\parsep}{0pt}
     \setlength{\leftmargin 12.7pt}{\rightmargin 0pt}
         \setlength{\itemsep}{0pt} \settowidth
    {\labelwidth}{#1.}\sloppy}}{\end{list}}

\newcounter{itemlistc}
\newcounter{romanlistc}
\newcounter{alphlistc}
\newcounter{arabiclistc}

\newcommand{\fcaption}[1]{
        \refstepcounter{figure}
        \setbox\@tempboxa = \hbox{\footnotesize Fig.~\thefigure. #1}
        \ifdim \wd\@tempboxa > 5in
           {\begin{center}
        \parbox{5in}{\footnotesize\smalllineskip Fig.~\thefigure. #1}
            \end{center}}
        \else
             {\begin{center}
             {\footnotesize Fig.~\thefigure. #1}
              \end{center}}
        \fi}

\newcommand{\tcaption}[1]{
        \refstepcounter{table}
        \setbox\@tempboxa = \hbox{\footnotesize Table~\thetable. #1}
        \ifdim \wd\@tempboxa > 5in
           {\begin{center}
        \parbox{5in}{\footnotesize\smalllineskip Table~\thetable. #1}
            \end{center}}
        \else
             {\begin{center}
             {\footnotesize Table~\thetable. #1}
              \end{center}}
        \fi}

\def\pmb#1{\setbox0=\hbox{#1}
    \kern-.025em\copy0\kern-\wd0
    \kern.05em\copy0\kern-\wd0
    \kern-.025em\raise.0433em\box0}

\def\fnt#1#2{\footnotetext{\kern-.3em
    {$^{\mbox{\scriptsize #1}}$}{#2}}}

\def\fpage#1{\begingroup
\voffset=.3in
\thispagestyle{empty}\begin{table}[b]\centerline{\footnotesize #1}
    \end{table}\endgroup}

\def\runninghead#1#2{\pagestyle{myheadings}
\markboth{{\protect\footnotesize\it{\quad #1}}\hfill}
{\hfill{\protect\footnotesize\it{#2\quad}}}}
\font\tenrm=cmr10
\font\tenit=cmti10
\font\tenbf=cmbx10
\font\bfit=cmbxti10 at 10pt
\font\ninerm=cmr9

\font\eightrm=cmr8

\def\FigName{figure}%
\newbox\captionbox
\long\def\@makecaption#1#2{%
  \ifx\FigName\@captype
    \vskip\abovecaptionskip
    \setbox\tempbox\hbox{{\figurecaptionfont #1\hskip1em #2}}
    \ifdim\wd\tempbox< 28pc
    \centerline{\box\tempbox}
    \else
    {\figurecaptionfont #1\hskip1em #2\par}
\fi\else
    \setbox\tempbox\hbox{{\tablecaptionfont #1\hskip1em #2}}
    \ifdim\wd\tempbox< 28pc
    \centerline{\box\tempbox}
    \else
    {\tablecaptionfont #1\hskip1em #2\par}%
    \fi
 \vskip\belowcaptionskip
 \fi}
\InputIfFileExists{psfig.sty}
{\typeout{^^Jpsfig.sty inputed...ok}}{\typeout{^^JWarning: psfig.sty could be be found.^^J}}
\InputIfFileExists{epsfsafe.tex}
{\typeout{^^Jepsfsafe.tex inputed...ok}}
            {\typeout{^^JWarning: epsfsafe.tex could not be found.^^J}}
\InputIfFileExists{epsfig.sty}
{\typeout{^^Jepsfig.sty inputed...ok}}{\typeout{^^JWarning: epsfig.sty could not be found.^^J}}
\InputIfFileExists{epsf.sty}
{\typeout{^^Jepsf.sty inputed...ok}}{\typeout{^^JWarning: epsf.sty could not be found.^^J}}%
\def\fps@figure{tbp}
\def\ftype@figure{1}
\def\ext@figure{lof}
\def\fnum@figure{Fig.\ \thefigure}
\def\qed{\hbox{${\vcenter{\vbox{              
   \hrule height 0.4pt\hbox{\vrule width 0.4pt height 6pt
   \kern5pt\vrule width 0.4pt}\hrule height 0.4pt}}}$}}

\begin{document}
\setlength{\textheight}{8.0truein}    

\runninghead{FULLY MULTI-QUBIT ENTANGLED STATES}
            {J.-M. Cai, Z.-W. Zhou and G.-C. Guo}

\normalsize\textlineskip
\thispagestyle{empty}
\setcounter{page}{1}

\copyrightheading{} 

\vspace*{0.88truein}

\fpage{1}
\centerline{\bf
FULLY MULTI-QUBIT ENTANGLED STATES} \vspace*{0.035truein}

\vspace*{0.37truein} \centerline{\footnotesize
JIAN-MING CAI\footnote{Email:jmcai@mail.ustc.edu.cn}}
\vspace*{0.015truein} \centerline{\footnotesize\it Key Laboratory of
Quantum Information, University of Science and Technology of China}
\baselineskip=10pt \centerline{\footnotesize\it Chinese Academy of
Sciences, Hefei, Anhui 230026, China} \vspace*{10pt}
\centerline{\footnotesize ZHENG-WEI ZHOU} \vspace*{0.015truein}
\centerline{\footnotesize\it Key Laboratory of Quantum Information,
University of Science and Technology of China} \baselineskip=10pt
\centerline{\footnotesize\it Chinese Academy of Sciences, Hefei,
Anhui 230026, China} \vspace*{10pt} \centerline{\footnotesize
GUANG-CAN GUO} \vspace*{0.015truein} \centerline{\footnotesize\it
Key Laboratory of Quantum Information, University of Science and
Technology of China} \baselineskip=10pt \centerline{\footnotesize\it
Chinese Academy of Sciences, Hefei, Anhui 230026, China}
\vspace*{0.225truein}

\vspace*{0.21truein}
\abstract{
We investigate the properties of different levels of entanglement in
graph states which correspond to connected graphs. Combining the
operational definition of graph states and the postulates of
entanglement measures, we prove that in connected graph states of
$N>3$ qubits there is no genuine three-qubit entanglement. For certain classes of graph states,
all genuine $k$-qubit entanglement, $2\leq k\leq N-1$, among every $k$ qubits vanishes.
These results about connected graph states naturally lead to the definition of fully multi-qubit
entangled states. We also find that the connected graph states of
four qubits is one but not the only one class of fully four-qubit
entangled states.}

\vspace*{10pt} \keywords{Graph States, Multi-qubit Entanglement,
Quantum Computation} \vspace*{3pt} 

\vspace*{1pt}\textlineskip  
\section{Introduction}           
\vspace*{-0.5pt} \noindent
The trend of quantum information processing is to implement large
scale quantum computation with many qubits \cite{DiVincenzo}. One
prospective proposal is the one-way quantum computation model, based
on some special kind of multi-particle entangled states and single
qubit measurements \cite{onewayQC}. The universal resource in
one-way quantum computer is the so called graph states that
correspond to mathematical graphs \cite{Clusterstate}, where the
vertices of the graph play the role of quantum spin systems and
edges represent Ising interactions. Graph states also have
applications in quantum communication of many users, e.g. open
destination quantum teleportation \cite{Pan}. Moreover, various
quantum error correcting codes for protecting quantum information
against decoherence are also graph states \cite{ECC}.

On the other hand, it is well known that entanglement is the most
fascinating feature of quantum mechanics. Very recently,
entanglement in interacting many-body systems becomes an increasing
important concept in condensed matter physics, such as quantum phase
transitions \cite{QPT}, superconductivity and fractional quantum
Hall effect \cite{cdp}. However, the structure and nature of
entanglement in multi-particle entangled states is not very clear
now. The most obstacle is that there is no known measure which can
completely characterize the entanglement of multi-particle entangled
states. Therefore, the study of entanglement properties of the
special significant multi-particle entangled states - graph states
is a very important and interesting topic
\cite{Dur,Hein,Hein2,Bell,Nest}.

In Refs.\cite{Hein}, Hein \textit{et al} characterize and quantify
the genuine multi-particle entanglement of graph states by the
Schmidt measure. They provide the upper and lower bounds of the
Schmidt measure \cite{Schmidt} in graph theoretical terms. In this
paper, we investigate the entanglement properties of graph states
from the viewpoint of different levels of entanglement. The main
result is that, using the operational description of graph states
and the fact that entanglement measures always decrease under local
operations and classical communications (LOCC), we present a simple
proof that in general connected graph states of $N>3$ qubits, there is
no genuine three-qubit entanglement. Moreover, for some classes of connected
graphs states, all genuine $k$-qubit entanglement, $2\leq k\leq N-1$,
among every $k$ qubits always vanishes. These results explicitly demonstrate
that graph states are indeed a kind of fully multi-qubit entangled states. In
addition, we find that the connected graph states of four qubits is
only one class of fully multi-qubit entangled states. We construct
different kinds of fully multi-qubit entangled states that are not
local unitary equivalent to connected graph states of four qubits.

\section{Graph states}
\noindent

Each mathematical (undirected, finite) \emph{graph} is denoted as \cite%
{graph}
\begin{equation}
G=(V,E)
\end{equation}%
where the finite set $V\subset \mathcal{N}$ is the set of vertices,
and the set $E\subset \lbrack V]^{2}$ is the set of edges. In the
context of graph states, people restrict to the simple graphs, which
contain neither edges connecting vertices with itself nor multiple
edges. Given a subset of vertices $S\subset V$, we can define the
\emph{subgraph }generated by $S$ as
$G_{S}=(S,E_{S})$, where $E_{S}$ $\subset E,$ and for every edges $%
\{a,b\}\in E$, if and only if $a,b\in S$ then $\{a,b\}\in E_{S}$.

For a given vertex $a\in V$, its \emph{neighborhood} $N_{a}\subset
V$ is defined as the set of vertices adjacent to the given vertex
$a$, i.e. the set of vertices $b\in V$ for which $\{a,b\}\in E$. For
two vertices $a,b\in V $, we say $a$ and $b$ is connected if there
exists an ordered list of
vertices $a=a_{1},a_{2},\cdots ,a_{n-1},a_{n}=b$ such that for all $i$, $%
(a_{i},a_{i+1})\in E$. If any two $a,b\in V$ are connected, the
graph is a \emph{connected graph}, otherwise it is a disconnected
graph which can be viewed as a collection of several separate
connected subgraphs.

Graph states that correspond to a mathematical graphs $G=(V,E)$ is a
certain pure quantum state on the Hilbert space
$H=(\mathcal{C}^{2})^{\otimes N}$, where $N=|V|$ is the number of
the vertices.
For every vertex $a\in V$ of the graph $G=(V,E)$, one can define a
Hermitian operator,
\begin{equation}
K_{G}^{a}=X_{a}\bigotimes\limits_{b\in N_{a}}Z_{b}
\end{equation}%
where the matrices $X_{a}$, $Y_{a}$ and $Z_{a}$ are Pauli matrices,
the lower index specifies the qubit on which the operators acts.
The \textit{graph state} $|G\rangle $ associated with the graph
$G=(V,E)$ is
the unique $n-$qubit state fulfilling%
\begin{equation}
K_{G}^{a}|G\rangle =|G\rangle
\end{equation}

The graph state $|G\rangle $ can be obtained by applying a sequence
of unitary two-qubit operations to the initial state $|+\rangle
^{\otimes N}$ as follows,
\begin{equation}
|G\rangle =\prod\limits_{\{a,b\}\in E}U_{ab}|+\rangle ^{\otimes N}
\end{equation}%
where $|+\rangle =(|0\rangle +|1\rangle )/\sqrt{2}$, and $|0\rangle
,|1\rangle $ are eigenvectors of $Z$ with eigenvalues $\pm 1$. The
unitary two-qubit operation $U_{ab}$ is a controlled $Z$ on qubits
$a$ and $b$, i.e. $U_{ab}=|00\rangle \langle 00|+|01\rangle \langle
01|+|10\rangle \langle 10|-|11\rangle \langle 11|$. We note that
theses unitary two-qubit operations commute with each other.
Therefore, we can adopt different orders of the sequence of $U_{ab}$
to the initial state $|+\rangle ^{\otimes N}$ and yield the same
graph state $|G\rangle $. This property is the key point in our
following proof.

As discussed above, a disconnected graph can be viewed as a
collection of several separate connected subgraphs. Therefore, a
disconnected graph state is just a product state of the
corresponding connected subgraph states. Without loss of generality,
it is sufficient for us to consider only the connected graph states
here. The entanglement structure in multi-qubit entangled states is
much more complex than the situation of two-qubit entangled states.
For pure states of $N$ qubits, there are different levels of genuine
$k$-qubit entanglement, $2\leq k\leq N$, which is shared among all
the $k$ qubits.

In Refs. \cite{Hein2}, Hein \textit{et al }show\textit{\ }that there is no $%
2 $-qubit entanglement between any two qubits in general $N$-vertex
connected graph states with $N\geq 3$ by examining the properties of
reduced
density matrices. However, there is no exact results about general genuine $%
k $-qubit entanglement for $3\leq k\leq N-1$. One reason is that unlike $2$%
-qubit entanglement entanglement \cite{Plenio}, there are few well
defined genuine multi-qubit entanglement measures
\cite{Mayer,Mintert,Localizable
entanglement,Barnum,Hyperdeterminants,Lamata,Lohmayer,ITMGE},
especially for general multi-qubit mixed states. For a natural
entanglement measure, it should satisfy several necessary
conditions, such as invariant under local unitary operations, vanish
for separable states, and decrease on average under LOCC. In the
following, using the operational definition of graph states and the
postulates of entanglement measures, we first prove that genuine
three-qubit entanglement vanish in connected graph states of four
qubits. Then we will generalize our results to arbitrary genuine
$k$-qubit entanglement.

\section{Genuine three-qubit entanglement in graph states}
\noindent In this section, we will investigate genuine three-qubit
entanglement among every three qubits in connected graph states.
Genuine three-qubit entanglement is the special kind of entanglement
which critically involves
all of the three qubits. For example, in GHZ state of three qubits $%
|GHZ\rangle =\frac{1}{\sqrt{2}}(|000\rangle +|111\rangle )$, the
state becomes separable if any one qubit is lost, which means that
the entanglement in GHZ state is the only global property shared by
all of the
three qubits. However, in W state \cite{Wstate} of three qubits $|W\rangle =%
\frac{1}{\sqrt{3}}(|001\rangle +|010\rangle +|100\rangle )$,
two-qubit entanglement will still exist when one qubit is neglected.
One well defined measure for genuine three-qubit entanglement is the
square root of CKW tangle \cite{CKW} proposed by Wootters
\textit{etc}, of which $\tau (|GHZ\rangle )=1$ and $\tau (|W\rangle
)=0$.

\begin{figure} [htbp]
\vspace*{13pt}
\centerline{\psfig{file=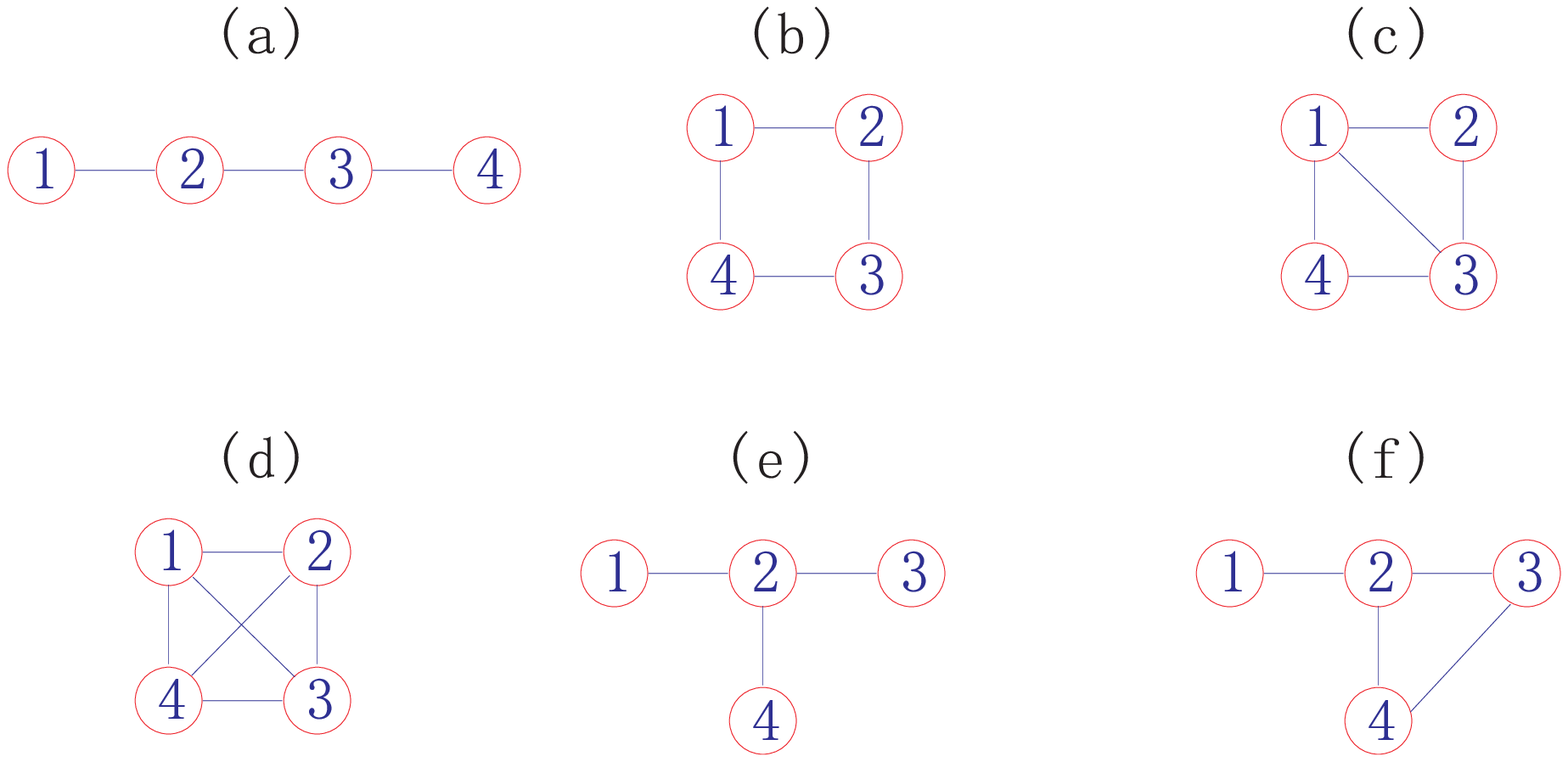, width=8.2cm}} 
\vspace*{13pt} \fcaption{Six classes of four vertices connected
graphs that are nonequivalent under graph isomorphisms.}
\end{figure}

\emph{Lemma 1: Genuine three-qubit entanglement vanishes in
connected graph states of four qubits.}

\emph{Proof:} There are six classes of four vertices connected
graphs that are nonequivalent under graph isomorphisms as depicted
in Fig. (1). For each
graph state $|G\rangle _{1234}$, we can write the reduced density matrix $%
\rho _{ijk}$ of every three qubit $i,j,k=1,2,3,4$. By exploiting
some skills, it is easy for us to construct a special pure states
decomposition of $\rho _{ijk}$ as $\rho _{ijk}=\sum\limits_{i}|\phi
_{i}^{\prime }\rangle \langle \phi _{i}^{\prime }|$, where $|\phi
_{i}^{\prime }\rangle$ are separable. For example, in the graph
state $|G_{d}\rangle _{1234}$ corresponding to Fig 1.(d), $\rho
_{123}=\frac{1}{2}(|\phi _{1}\rangle \langle \phi _{1}|+|\phi
_{2}\rangle \langle \phi _{2}|)$, where $|\phi _{1}\rangle
=\frac{1}{2}(|+00\rangle +|-01\rangle +|-10\rangle -|+11\rangle ) $
and $|\phi _{2}\rangle =\frac{1}{2}(|-00\rangle -|+01\rangle
-|+10\rangle
-|-11\rangle )$. The special pure states decomposition for $\rho _{123}$ is $%
\rho _{123}=\sum\limits_{i=1}^{_{{2}}}|\phi _{i}^{\prime }\rangle
\langle \phi _{i}^{\prime }|$, where $(|\phi _{1}^{\prime }\rangle
,|\phi
_{2}^{\prime }\rangle )^{T}=U(|\phi _{1}\rangle ,|\phi _{2}\rangle )$, and $%
U=\frac{1}{\sqrt{2}}\left(
\begin{array}{cc}
1 & i \\
1 & -i%
\end{array}%
\right) $. It is obvious that $|\phi _{1}^{\prime }\rangle $ and
$|\phi _{2}^{\prime }\rangle $ are separable for bipartition $1|23$.
Actually, we could show that the square root of CKW tangle $\tau
(\rho _{123})=0$ in a similar way. Therefore we conclude that
genuine three-qubit entanglement vanishes for every three
qubits.$\Box$

It should be emphasized that we only need to consider 2 equivalence
classes under local Clifford (LC) operations \cite{Hein2}. One class
includes Fig
1.(a), (b), (c), and (f), the other class includes Fig 1. (d) and (e) \cite%
{Nest2}. Based on lemma 1, we present the following theorem about
genuine three-qubit entanglement in general connected graph states
of more than three qubits.

\emph{Theorem 1: There is no genuine three-qubit entanglement in
general connected graph states of more than three qubits.}

\emph{Proof:} We examine genuine three-qubit entanglement among
every three qubits. Without loss of generality, we can denote these
three qubits as $1$,
$2$, and $3$. According to whether the subgraph $G_{\{1,2,3\}}=(\{1,2,3%
\},E_{\{1,2,3\}})$ is connected, there are two kinds of situations.

(a). The subgraph $G_{\{123\}}$ is connected. Since the number of qubits $%
N>3 $ and the corresponding graph is connected, there must exist one
qubit $4 $ which make the subgraph
$G_{\{1,2,3,4\}}=(\{1,2,3,4\},E_{\{1,2,3,4\}})$ is also connected,
i.e. it is one class of graph depicted in Fig 1. (a1). In
the first step to obtain the graph state $|G\rangle $, we get the state $%
|\psi \rangle =\prod\limits_{\{i,j\}\in
E_{\{1,2,3,4\}}}U_{ij}|+\rangle ^{\otimes N}$, i.e. $|\psi \rangle $
=$|G_{\{1,2,3,4\}}\rangle |+\rangle ^{\otimes N-4}$, where
$|G_{\{1,2,3,4\}}\rangle $ is the connected graph states of four
qubits, which leads to that genuine three-qubit entanglement in
$\rho _{123}$ is $0$ for $|\psi \rangle $ according to lemma 1.
(a2). In the second step, we apply unitary two-qubit operations
related to qubit $1$, $2$ and $3$ to $\psi \rangle $ and obtain
$|\psi ^{\prime }\rangle =\prod\limits_{i\in V-\{1,2,3,4\},j\in
\{1,2,3\},\{i,j\}\in {E}}U_{ij}|\psi \rangle $. The effects of each
block of operations $\prod\limits_{j\in \{1,2,3\},\{i,j\}\in
{E}}U_{ij}$ for $i\in V-\{1,2,3,4\}$ on $\rho _{123}$
can be characterize by the superoperator $\varepsilon (\rho _{123})=\frac{1}{%
2}\rho _{123}+\frac{1}{2}(u_{1}\otimes u_{2}\otimes u_{3})\rho
_{123}(u_{1}\otimes u_{2}\otimes u_{3})^{\dag }$, where $u_{1}$, $u_{2}$, $%
u_{3}=I_{2}$ or $Z$. It describes a certain local operation on qubit
$1$, $2$ and $3$. Since any entanglement measure should be an
entanglement monotone function \cite{Plenio}, thus genuine
three-qubit entanglement in $\rho _{123} $ is $0$ for $|\psi
^{\prime }\rangle $. (a3). In the last step, we apply the remain
unitary two-qubit operations independent on qubit $1$, $2$
and $3$ to $|\psi ^{\prime }\rangle $ and obtain the final graph state $%
|G\rangle =\prod\limits_{i,j\in V-\{1,2,3\},\{i,j\}\in
{E}}U_{ij}|\psi ^{\prime }\rangle $. In this step, $\rho _{123}$ is
unchanged. Therefore, we conclude that genuine three-qubit
entanglement in $\rho _{123}$ vanishes for $|G\rangle $.

(b) The subgraph $G_{\{1,2,3\}}$ is disconnected. In this situation,
we first get $|\psi \rangle =\prod\limits_{\{i,j\}\in
E_{\{1,2,3\}}}\\ U_{ij}|+\rangle ^{\otimes N}$. Since
$G_{\{1,2,3\}}$ is disconnected, it is obvious that the state $|\psi
\rangle $ is separable. The following steps are similar to the above
situation (a). Therefore, we also obtain that genuine three-qubit
entanglement in $\rho _{123}$ vanishes for the state $|G\rangle $,
and theorem 1 is proved. $\Box $

Our results imply that for general connected graph states of more
than three qubits, if we consider the reduced states of every three
qubits by tracing out the other qubits, no genuine three-qubit
entanglement exists. In other words, entanglement in these graph
states is the properties that critically involves more than three
qubits which is similar to the case of GHZ states.

\section{General genuine k-qubit entanglement in graph states}

To our knowledge, the problem about general multipartite entanglement is extremely hard and remain far from clear, in particular there is no well-defined general genuine multi-qubit entanglement measures. Thus, it is very difficult to investigate the properties of general genuine $k$-qubit entanglement in graph states. Even though, for some classes of
graph states, it is still possible for us to tackle this problem.

\emph{Theorem 2: Consider a connected graph states of $N$ qubits, and let $S$ be a group of
$m$ qubits, $2\leq m\leq N-1$, the reduced density matrix $\rho _{S}$ is separable for certain bipartition
if the subgraph $G[S]$ can be made disconnect by using all possible local complementations of the graph.}

{\it Proof:} It is known that the equivalence of graphs under local
complementation implies the local Clifford equivalence of the
corresponding graph states with the same entanglement properties
\cite{Nest2004}. Thus, if the subgraph $G[S]$ can be made disconnect
by using local complementations of graph, the reduced density matrix
$\rho_{S}$ is local unitary equivalent to an ensemble of the graph
state $|G[S]\rangle$ with possible local Pauli $Z$ operations.
Therefore, $\rho _{S}$ is bipartite separable.

From the above theorem 2, it is easy for us to make the statement
about the properties of all genuine $k$-qubit entanglement for some
classes of connected graph states.

\emph{Corollary 1: All genuine $k$-qubit entanglement, $2\leq k\leq
N-1,$ vanishes in connected graph states of $N$qubits which are
locally equivalent to $1D$ cluster states, complete graph and tree
graph states.}

\emph{Proof: } According to theorem 2, for these graph states which
are locally equivalent to $1D$ cluster states, complete graph and
tree graph states, given any subset $S$ of $2\leq k\leq N-1$ qubits,
the reduced density matrix $\rho _{S}$ is separable for some
bipartition, therefore any genuine $k$-qubit entanglement measure
should be zero, which is one of the necessary conditions for
entanglement measures. Thus we finish the proof of corollary 1.

The definition of genuine multipartite entanglement here is based on
the idea of residual entanglement \cite{CKW}, which is different
from others, e.g. entanglement measures based on GHZ extraction
yield and in stabilizer formalism \cite{David04, Bravyi06}. Whether
the above corollary 1 is valid for general graph states remains an
interesting open problem.

\section{Fully multi-qubit entangled states}

The connected graph states can not be written as a product form for
any bipartition, and it is believed that connected graph states are
genuine multi-qubit entangled states, which is supported by our result in
corollary 1. Moreover, it is easy to check that the reduced
density matrix of each qubit is $I/2$. With these intuitions, we
could naturally define the fully multi-qubit entangled states as

\textit{A pure state of $N$ qubits $\left\vert \psi \right\rangle $
is fully $N$-qubit entangled if it satisfies: (1) There does not
exist a bipartition such that $\left\vert \psi \right\rangle $ is
product; (2) The reduced state of each qubit is maximally mixed,
i.e. $\rho _{i}=Tr_{1,2,\cdots ,i-1,i+1,\cdots N}\left\vert \psi
\right\rangle \langle \psi |=I/2$; (3) There is no genuine $k$-qubit
entanglement for $2\leq k\leq N-1$.}

If $N=2$, the above definition will reduce to maximally two-qubit
entangled states. We note that stronger and slight different
definitions of maximal multipartite entanglement appeared in Refs
\cite{Gisin,MED}. However, the criterions we propose above are from
the viewpoint of different levels of entanglement in multi-qubit
entangled states, and is strongly motivated by the important class
of multi-qubit entangled states, i.e. graph states. The above
condition 1 means that $\left\vert \psi \right\rangle $ is not
separable. The condition 2 stems from the complementary relations in
multi-qubit entangled states \cite{Horodecki,ICL}, the fact that the
reduced state of each qubit is maximally mixed implies that local
information is minimum, i.e. maximum entanglement. The last
condition is introduced according to different levels of
entanglement structure in multi-qubit entangled states. For example,
in general $N$-qubit GHZ states there is only genuine $N$-qubit
entanglement, which is shared among all of the $N$ qubits. However,
in general $N$-qubit W states, there are only two-qubit
entanglement, i.e. shared only between pairs of qubits. In this sense, $N$%
-qubit GHZ states are fully $N$-qubit entangled states, while $N$%
-qubit W states are not.

The connected graph states of four qubits $|G_{4}\rangle $ is a kind
of fully four-qubit entangled states. However, it not the only class
of fully four-qubit entangled states. A generic pure state of four
qubits can always be transformed to the normal form state by the
determinant 1 SLOCC (stochastic local operations and classical
communication) operations \cite{Verstraete2002,Normal form},
$G_{abcd}=\frac{a+d}{2}(|0000\rangle
+|1111\rangle )+\frac{a-d}{2}(|0011\rangle +|1100\rangle )+\frac{b+c}{2}%
(|0101\rangle +|1010\rangle )+\frac{b-c}{2}(|0110\rangle
+|1001\rangle )$, where $a,b,c,d$ are complex parameters with
nonnegative real part. Without loss of generality, we could assume
$A=(a+d)/2$ is a positive real number,
i.e. $A=x_{1}=|A|$. We denote $B=(b+c)/2{=}x_{2}\exp {(i\phi }_{2}{)}%
,C=(a-d)/2=x_{3}\exp {(i\phi }_{3}{)},D=(b-c)/2{=}x_{4}\exp {(i\phi }_{4}{)}$%
, with $x_{2}=|B|$, $x_{3}=|C|$, $x_{4}=|D|$. We consider those
$G_{abcd}$
that is a non-product state. The reduced density matrix of each qubit in $%
G_{abcd}$ is $I/2$. The square root of CKW tangle of the mixed
states
obtained by tracing out one qubit of $G_{abcd}$ is always equal to zero \cite%
{Verstraete2002}. To ensure that pairwise entanglement also vanish in $%
G_{abcd}$, the parameters should fulfill the following conditions

\begin{eqnarray}
2x_{1}x_{2}|\cos {\phi }_{2}| &\leq &x_{3}^{2}+x_{4}^{2}  \nonumber \\
2x_{3}x_{4}|\cos ({\phi }_{3}-{\phi }_{4})| &\leq
&x_{1}^{2}+x_{2}^{2}
\nonumber \\
2x_{1}x_{3}|\cos {\phi }_{3}| &\leq &x_{2}^{2}+x_{4}^{2}  \nonumber \\
2x_{2}x_{4}|\cos ({\phi }_{2}-{\phi }_{4})| &\leq &x_{1}{}^{2}+x_{3}^{2} \\
2x_{1}x_{4}|\cos {\phi }_{4}| &\leq &x_{2}^{2}+x_{3}{}^{2}  \nonumber \\
2x_{2}x_{3}|\cos ({\phi }_{2}-{\phi }_{3})| &\leq
&x_{1}^{2}+x_{4}^{2} \nonumber
\end{eqnarray}%
Therefore, we can easily construct an explicit example of fully
four-qubit entangled states

\begin{equation}
\left\vert MG_{4}\right\rangle =c(|0000\rangle +|1111\rangle )+i\sqrt{\frac{1%
}{2}-c^{2}}(|0011\rangle +|1100\rangle )
\end{equation}%
where $0\leq c\leq \sqrt{1/2}$. It is obvious that $\left\vert
MG_{4}\right\rangle $ is not always local unitary equivalent to any
four-qubit connected graph state $|G_{4}\rangle $. This can be
verified by noting that $Tr\rho _{23}^{2}$ is not always the same
for $\left\vert MG_{4}\right\rangle $ and $\left\vert
G_{4}\right\rangle $, where $\rho _{23} $ is the reduced density
matrix of qubit $2$ and $3$.

\section{Conclusions and Discussions}
\noindent
In conclusion, we have investigate the entanglement
properties of graph states by calculating different levels of
genuine $k$-qubit entanglement. In this paper, via the operational
definition of graph states, and using the postulates of entanglement
measures, we construct explicit proofs that there are no genuine
three-qubit entanglement in general connected graph states of $N>3$ qubits.
For some class of graph states, all $k$-qubit entanglement vanishes for
$2\leq k\leq N-1$. These results, together with intuitions
about graph states, lead to a definition of fully multi-qubit
entangled states. In addition, we find that the connected graph
states of four qubits is only one class of fully multi-qubit
entangled states. We present a kind of fully multi-qubit entangled
states that are not always local unitary equivalent to connected
graph states of four qubits. Our results demonstrate exactly that
graph states are genuine multi-qubit entangled states. It may help
us to gain some insight into the complex entanglement structure of
multi-qubit entangled states from a new viewpoint of different
levels of entanglement.

\nonumsection{Acknowledgements} \noindent We gratefully thanks
Wolfgang D\"{u}r, Li Yu, Andreas Osterloh, Jens Siewert, and Lucas Lamata
for valuable suggestions and helpful discussions. This work was
funded by National Fundamental Research Program, the Innovation
funds from Chinese Academy of Sciences, NCET-04-0587, and National
Natural Science Foundation of China (Grant No. 60121503, 10574126).

\nonumsection{References}

\end{document}